\documentstyle[12pt]{article}

\def\beq{\begin{equation}}
\def\eeq{\end{equation}}
\def\bea{\begin{eqnarray}}
\def\eea{\end{eqnarray}}
\def\nn{\nonumber}

\begin{document}

\begin{titlepage}
\begin{center}
Nov.'91 \hfill  TIT/HEP-181\\
{\large \bf Puncture Operator in c=1 Liouville Gravity}\\[.4in]
\bf Y.Kitazawa \\
\it Department of Physics\\
Tokyo Institude of Technology\\
Oh-Okayama, Meguro-ku, Tokyo 152, Japan

\end{center}

\vskip.3in
\vspace{\fill}
\centerline{\bf ABSTRACT}
\begin{quotation}
We identify the puncture operator in $c=1$ Liouville gravity as
the discrete state with spin $J=\frac{1}{2}$.
The correlation functions involving this operator satisfy
the recursion relation which is characteristic in topological
gravity. We derive the recursion relation involving the
puncture operator by the operator product expansion.
Multiple point correlation functions are determined recursively
from fewer point functions by
this recursion relation.

\end{quotation}
\end{titlepage}
\newpage

\newpage
Through recent progress in our understanding of 2 dimensional
gravity by matrix models\cite{matrix}, we have gained precious insight
into universal properties of quantum gravity.
It becomes clear that the topological gravity point of view
advocated by Witten\cite{witten} goes quite far to capture the essential
features of the 2d gravity. In fact the most remarkable aspects
of the 2d gravity, namely the KdV structure\cite{douglas} and
Virasoro like constraints\cite{kawai}\cite{verlinde}
represent such topological aspects of the theory.

However we should also recognize that there are also geometrical
aspects of the theory such as intrinsic and extrinsic fractal
dimensions which is beyond the scope of the topological gravity.
The Liouville theory approach to the 2d gravity and string theory
has the advantage that it can capture both topological and geometrical
aspects of the theory.

One of significant recent developments in Liouville gravity is
the successful
evaluation of correlation functions\cite{kitazawa}.
Certainly Liouville theory is more awkward to capture the topological
aspects of the 2d gravity than the powerful matrix models.
Nevertheless it is important to push the Liouville theory approach
to see how far it can go to understand topological aspects of the
theory from the continuum theory point of view.
2d gravity is our important foothold to study higher dimensional
quantum gravity and general string theory.
The significance of deeper understanding of Liouville gravity
stems from it.
We hope that such investigation is useful to make further progress
in both more general quantum gravity theory and string theory.

The discrete states in $c=1$ 2d Liouville gravity
were interpreted as topological
states in \cite{polyakov} and they were further studied in
\cite{sakai}.
Recently Witten\cite{witten1},Klebanov and Polyakov\cite{kleb}
have shown that the
discrete states in $c=1$ 2d gravity form the operator algebra which
is isomorphic to the area preserving diffeomorphism of the plane.
The holomorphic part of the discrete states are
\beq
\Psi_{J,m} (z) = \psi_{J,m} (z) exp(\sqrt{2}(-1+J)\phi (z)),
\eeq
where $\psi _{J,m}(z)$ are well-known $SU(2)$ multiplets in $c=1$
conformal field theory.
After rescaling the operators, they are found to obey
\beq
\Psi_{J_1,m_1}(z)\Psi_{J_2,m_2}(0) = z^{-1}(J_2m_1 - J_1m_2)
\Psi_{J_1+J_2-1,m_1+m_2}(0).
\eeq
In order to make physical operators in the closed string theory,
we need to combine the holomorphic and antiholomorphic sectors.
Therefore the observables are
\beq
V_{J,m,m'} = \int d^2z\Psi_{J,m}(z)\bar{\Psi}_{J,m'}(\bar{z}).
\eeq
The theory poseses $SU(2)_L \times SU(2)_R$ symmetry.

Although Polyakov and Klebanov have studied most general `resonant'
three point functions of discrete states,
it has not been made clear what principle determines general $n$ point
amplitudes.
The `resonant' correlators conserve the Liouville charge.
They are considered to be `bulk' amplitudes if we regard
the Liouville field as an extra dimension.
Discrete states alone form closed operator algebra. It is consistent
to consider only discrete states and put aside tachyons
with general momentum. Furthermore we do not perturb the theory
with the cosmological constant operator.
This is the strategy
what we shall adopt in this paper.
The topological aspects of the theory may become more transparent
in this truncation of the theory.

Since the discete states are thought to be topological degrees of freedom,
their correlation functions are expected to obey recursion relations
which appear in general $c < 1$ matrix models and topological gravity.
The essential feature of these topological degrees of freedom
(gravitational descendents) can be understood if we consider the
topological gravity of Witten. In term of the Liouville gravity,
it is $c=-2$ matter coupled to 2d gravity.

In this theory, there are infinitely many operators $\sigma _i$
with $i=0,1,2,\cdots$. The connected Green's functions on the
sphere are
\beq
<\sigma _{i_1}\cdots\sigma _{i_n}PPP> = n!,
\eeq
where $P=\sigma _0$ and $\sum _{j=1}^n ( i_j -1) =0$ due to the
resonant condition.
We consider the generating function of the connected Green's functions.
Let us define a field
\beq
\Phi(\phi)=t_0e^{-\phi} + t_1 + t_2 e^{\phi} + \cdots,
\eeq
where $\phi$ represents Liouville coordinate.
The generating function is
\bea
W = \int d\phi (1 + \Phi + \Phi ^2 + \cdots )\nn\\
=\int d\phi \frac{1}{1-\Phi} ,
\eea
where $\phi$ integral imposes Liouville charge conservation.
The connected Green's functions are given by
\bea
<\sigma_{i_1}\sigma_{i_2}\cdots\sigma_{i_n}PPP>\nn\\
=
\frac{\partial}{\partial t_{i_1}}
\frac{\partial}{\partial t_{i_2}}
\cdots
\frac{\partial}{\partial t_{i_n}} W|_{t_i=0}.
\eea
The generating function neatly summarizes the symmetry of the
theory associated with the target space. The target space
in this case is the Liouville dimension.
In what follows we demonstrate that the Virasoro constraints
of this theory
can be interpreted as diffeomorphism invariance of the Liouville
dimension.

It is convenient to introduce a new variable $z=e^{\phi}$.
Let us consider the following infinitesimal
change of the integration variable
\beq
z \rightarrow z+cz^n.
\eeq
Then $W$ changes as follows
\beq
\int \frac{dz}{z} \frac{1}{1-\Phi - c((n-1)z^{n-1}+(z^n\partial_{z}
-(n-1)z^{n-1})\Phi}.
\eeq
Therefore the change of the integration variable is equivalent to
\bea
\Phi \rightarrow \Phi + \delta \Phi\nn\\
\delta \Phi = (n-1)z^{n-1} + \sum_{i}(i-n)t_iz^{i+n-1}.
\eea
It is interpreted as the change of the coupling constants $\{t_i\}$.
Since the generating function is invariant under the change of the
integration variable, the theory must be invariant under the change
of the coupling constants.
This requirement leads to the constraints on the correlation functions.
\beq
L_{n}<\sigma _{i_1}\cdots \sigma _{i_n}PPP> = 0,
\eeq
where $L_n = n\frac{\partial}{\partial t^{n+1}} + \sum _i (i-n-1)t_i
\frac{\partial}{\partial t^{i+n}}$ with $n \geq -1$.
$L_n$ operators form the Virasoro algebra
\beq
[L_n,L_m] = (n-m) L_{n+m}.
\eeq
In particular
\beq
L_{-1} = -\frac{\partial}{\partial t_0} + \sum_{i} i t_i
\frac{\partial}{\partial t^{i-1}}.
\eeq
$L_{-1}$ constraint leads to the following recursion relation
\bea
<\sigma _{0}\sigma _{i_1} \cdots \sigma _{i_n}PPP>\nn\\
=\sum_{k=1}^{n}<i_{k} \sigma _{i_k-1}\sigma _{i_1}\cdots
\hat{\sigma_{i_k}}\cdots \sigma _{i_n}PPP>,
\eea
where $\hat{\sigma}$ implies the absence of such an operator.
This is the famous recursion relation of Witten\cite{witten}.
It is strong enough to determine all correlation functions on the
sphere with just single input $<PPP> =1$.

Next we would like to pose analogous questions concerning the discrete
states in the $c=1$ Liouville gravity.
We hope to know some principle which relates different amplitudes
which contain different numbers of discrete states.
In particular we hope to identify the operator which acts like the
puncture operator $P$ in topological gravity.

Although these aspects have been investigated in
\cite{polch}\cite{polyakov}, our
understanding is far from complete. Especially little is known
concerning the puncture operator. In $(p,q)$ 2d gravity,
the operator with gravitational scaling
exponent(Liouville charge)
$\beta = -\sqrt{\frac{q}{2p}}$ appears to play the role
of the puncture operator.
It is a so-called mod $p$
or missing state. Therefore one might expect that in the
$c \rightarrow 1$
limit, it becomes a discrete state with isospin $J=\frac{1}{2}$.
Indeed we find that the operator with $J=\frac{1}{2}$ acts like
the puncture operator and it gives rise to the recursion relations
among the amplitudes.

Since $(m,m')$ dependence of the correlation functions are
governed by $SU(2)_L\times SU(2)_R$ symmetry,
we confine our attention to tachyonic discrete states,
$m=m'=\pm J$ and denote them by $V_{J,\pm J}$.
Since these states constitute $SU(2)$ multiplets with general
discrete states, it is natural to consider them as part of
discrete states.
The recursion relation we have found is
\bea
<\tilde{V}_{\frac{1}{2},\frac{1}{2}}
\tilde{V}_{J_1,- J_1}\cdots
\tilde{V}_{J_n,- J_n}
\tilde{V}_{J'_1,J'_1}\cdots
\tilde{V}_{J'_{n'},J'_{n'}}>\nn\\
= \sum_{i=1}^n (2J_i-1)
<\cdots
\tilde{V}_{J_i-\frac{1}{2},-J_i + \frac{1}{2}}
\cdots
\tilde{V}_{J'_1,J'_1}\cdots
\tilde{V}_{J'_{n'},J'_{n'}}>,
\label{recur}
\eea
where $\tilde{V}_{J,\pm J}$ are rescaled operators
\beq
V_{J,\pm J} = \pi \frac{\Gamma (1-2J)}{\Gamma (2J)}
\tilde{V}_{J,\pm J}.
\label{scale}
\eeq
Needless to say, analogous recursion relation holds with the opposite
sign of the $J_z'$s.

Since the correlation functions of the discrete states are singular,
we need to rescale the operators to obtain finite correlation
functions. In order to make the correlation functions well defined,
we adopt infinitesimal shift of the momentum,
namely $J \rightarrow J-\epsilon$.
In order to prove this recursion relation, we consider the following
operator product expansion(OPE).
\bea
e^{i\sqrt{2}(\frac{1}{2}-\epsilon )\varphi (z)}
e^{\sqrt{2}(-\frac{1}{2}-\epsilon ) \phi (z)}
\times
e^{-i\sqrt{2}J\varphi (0)}
e^{\sqrt{2}(J-1) \phi (0)}\nn\\
= z^{-2J(\frac{1}{2}-\epsilon )}z^{2(J-1)(\frac{1}{2}+\epsilon)}\nn\\
\times e^{i\sqrt{2}(\frac{1}{2}-\epsilon )\varphi (z)
-i\sqrt{2}J\varphi (0)}
e^{\sqrt{2}(-\frac{1}{2}-\epsilon ) \phi (z)
+\sqrt{2}(J-1) \phi (0)}\nn\\
= z^{-1+2\epsilon (2J-1)} e^{i\sqrt{2}(-J+\frac{1}{2})\varphi (0)}
e^{\sqrt{2}(J-\frac{1}{2}-1)\phi (0)}
+ \cdots.
\eea
Therefore
\bea
<V_{\frac{1}{2}-\epsilon,\frac{1}{2}-\epsilon}
V_{J,-J}\cdots >\nn\\
= \int dz^2 |z|^{-2+4\epsilon (2J - 1)}
<V_{J-\frac{1}{2},-J+\frac{1}{2}}\cdots > \nn\\
= 2\pi \int_0^{\infty} dr r^{-1+4\epsilon (2J -1)} e^{-\mu r}
<V_{J-\frac{1}{2},-J+\frac{1}{2}}\cdots > \nn\\
= 2\pi (\mu )^{-4\epsilon (2J_1 -1)} \Gamma (4\epsilon (2J_1-1))
<V_{J-\frac{1}{2},-J+\frac{1}{2}}\cdots > \nn\\
= \frac{\pi}{2\epsilon (2J-1)}
<V_{J-\frac{1}{2},-J+\frac{1}{2}}\cdots > ,
\eea
where we introduced another infinitesimal regularization parameter
$\mu$.
Let us rescale the operator as eq.(\ref{scale}).
If we note that
\bea
\frac{\Gamma (1-2(\frac{1}{2}-\epsilon ))}
{\Gamma (2(\frac{1}{2} -\epsilon ))} = \frac{1}{2\epsilon} \nn\\
\frac{\Gamma (1-2J)}{\Gamma (2J)} = \frac{1}{(2J-1)^2}
\frac{\Gamma (1-2(J-\frac{1}{2}))}{\Gamma (2(J-\frac{1}{2}))},
\eea
we obtain the recursion relation eq.(\ref{recur}).

Let us check our recursion relation by known results.
The $n+1$ point tachyon amplitudes with one negative
chirality state insertion is
\beq
<V_{\frac{n-1}{2},-\frac{n-1}{2}}V_{J_1,J_1}\cdots V_{J_n,J_n}>
= (n-2)!.
\label{cort}
\eeq
The resonant condition fixes the negative chirality states to be
$J=\frac{n-1}{2},m=-\frac{n-1}{2}$. It leads to another condition
$\sum_{i=1}^n J_i = \frac{n-1}{2}$. In order to satisfy
the latter condition, we need at least
one insertion of the cosmological constant operator $V_{0,0}$
in the correlator.
Let $J_1=\frac{1}{2}$, then the recursion relation (\ref{recur})
implies
\bea
<\tilde{V}_{\frac{n-1}{2},-\frac{n-1}{2}}\tilde{V}_
{\frac{1}{2},\frac{1}{2}}
\tilde{V}_{J_2,J_2}\cdots \tilde{V}_{J_n,J_n}> \nn\\
=(n-2)<\tilde{V}_{\frac{n-2}{2},-\frac{n-2}{2}}\tilde{V}_{J_2,J_2}
\cdots \tilde{V}_{J_n,J_n}> ,
\eea
which agrees with eq.(\ref{cort}).

How about the correlation functions which involve more
than single negative chirality states.
Such amplitudes with generic momenta are not singular
unlike the amplitudes with single negative chirality
state. If we insist on factorizing a pole due to the Liouville
charge conservation, the residue vanishes for the former unlike the
latter\cite{polyakov}.
However if we adopt the rescaling procedure
eq.(\ref{scale}),
the correlation functions are nonvanishing in both cases.
Such amplitudes have been investigated in \cite{jev}\cite{kuta}.
They allow the physical interpretation in terms of the
effective Lagrangian of massless scalar field(tachyon)
by introducing 1PI vertices.
Therefore such correlation functions are rather complicated.
1PI vertices could arise if we integrate
out discrete states and consider tachyon field only.
The strategy we adopt in this paper is rather different.
We consider the whole
discrete states and look for recursion relations
among the correlators.
We have checked that the general
resonant correlators of tachyonic
discrete states satisfy
the recursion relation eq.(\ref{recur}) up to 6 point functions.

Since we have rescaled the discrete states, all correlation
functions become finite. Before rescaling, the degree of the
divergence of a $n$ point function with $n_0$ insertions of
$V_{0,0}$ operator is $n-2n_0$.
The only resonant three point function of the discrete states is
\beq
<\tilde{V}_{\frac{1}{2},-\frac{1}{2}}\tilde{V}
_{\frac{1}{2},\frac{1}{2}}
\tilde{V}_{0,0}> = 1.
\eeq
The degree of the divergence of this correlator is 1.
The simple pole is due to the Liouville charge conservation.
For $n$ point functions, the maximum possible degree of divergence is
$n-2$.
This is because $n-3$ extra pole could arise due to on shell
intermediate states.
It explains why need at least single insertion of $V_{0,0}$ operator
for the rescaled correlation functions to be nonvanishing.

The recursion relation eq.(\ref{recur}) imposes very strong constraints on
the correlation functions. Let us consider the correlation functions
with single insertion of $V_{0,0}$.
The Liouville charge conservation requirement is
$ \sum _{i} (J_i-1) = -2 $. Let us take aside one $V_{0}$ operator
and two $V_{\frac{1}{2}}$ operators in the correlator.
For each operator with $J > 1$, we need $2J-2$
numbers of $V_{\frac{1}{2}}$
operators to balance the Liouville charge.
Therefore the relevant correlator is
\beq
<\tilde{V}_{J_1,-J_1}\cdots \tilde{V}_{J_n,-J_n}
\tilde{V}_{J'_1,J'_1}\cdots \tilde{V}_{J'_{n'},J'_{n'}}
\tilde{V}_{\frac{1}{2}}\cdots
\tilde{V}_{\frac{1}{2}}
\tilde{V}_{\frac{1}{2}}
\tilde{V}_{0}> .
\label{corm}
\eeq
where $\tilde{V}_{\frac{1}{2}}\cdots$ imply
$\sum _{i=1}^{n+n'} (2J_i-2)$ insertions of $\tilde{V}_
{\frac{1}{2}}$ .
Let us assume that $N$ point functions are known. Then by the
mathematical induction $N+1$ functions are determined by the
recursion relation eq.(\ref{recur}), since there always exist
$V_{\frac{1}{2}}$ operators in the correlation functions
we are considering.
By using the recursion relations the correlation function
eq.(\ref{corm}) is
determined to be
\beq
(\sum _{i=1}^n (2J_i-1))!(\sum _{j=1}^{n'} (2J'_j-1))!.
\eeq

For general discerete states we can write down the analogous
recursion relation
\bea
<\tilde{V}_{\frac{1}{2},\pm\frac{1}{2},\pm\frac{1}{2}}
\tilde{V}_{J_1,m_1,m'_1}\cdots
\tilde{V}_{J_n,m_n,m_n'}>\nn\\
= \sum_{i=1}^n (2J_1-1) \nn\\
C(\frac{1}{2},
\pm\frac{1}{2},
J_i,
m_i|J_i-\frac{1}{2},m_i\pm\frac{1}{2})
C(\frac{1}{2},
\pm\frac{1}{2},
J_i,m'_i|J_i-\frac{1}{2},m'_1\pm\frac{1}{2})
\nn\\
\times <\tilde{V}_{J_1,m_{1},m'_1}\cdots
\tilde{V}_{J_i-\frac{1}{2},m_i\pm \frac{1}{2},m'_i\pm \frac{1}{2}}
\cdots\tilde{V}_{J_n,m_n,m'_{n}}>,
\eea
where $
C(\frac{1}{2},
\pm\frac{1}{2},
J_i,m_i|J_i-\frac{1}{2},m_i\pm\frac{1}{2})$
is the Crebsh-Gordan coefficients.
Let us consider $V_{J,J-1}$ states which form Virasoro like closed
operator algebra for example.
The correlation functions of these states obey the following
recursion relation.
\bea
<\tilde{V}_{\frac{1}{2},-\frac{1}{2}}
\tilde{V}_{J_1,J_1-1}\cdots
\tilde{V}_{J_n,J_n-1}
\tilde{V}_{\frac{1}{2},\frac{1}{2}}
\tilde{V}_{0,0}
>\nn\\
= \sum_{i=1}^n \frac{(2J_i-1)^2}{2J_i}
<\cdots
\tilde{V}_{J_i-\frac{1}{2},J_i - \frac{3}{2}}
\cdots
\tilde{V}_{\frac{1}{2},\frac{1}{2}}
\tilde{V}_{0,0}
>.
\eea
By this recursion relation, the correlation functions of
these operators are found to be
\bea
<
\tilde{V}_{J_1,J_1-1}\cdots
\tilde{V}_{J_n,J_n-1}
\tilde{V}_{\frac{1}{2},\frac{1}{2}}
\tilde{V}_{0,0}>= (\prod _{i=1}^n ({\frac{1}{2J_i}})) (n-1)!
\eea
which is of the very familiar type.

In this paper we investigated the algebraic relations
among correlation functions of the discrete states in $c=1$ Liouville
gravity. They are found to obey the recursion relation which is
characteristic in topological gravity.
Certainly it is still just a begining. In order to make
the full contact with the recursion relations which are found in the
matrix model, we need to relate amplitudes of different genus.
We also need to derive other recursion relations which form
$W_{\infty}$ algebra. We would like to report progress in these
aspects in the near future.

{}~

{}~

\noindent{\large\bf Acknoledgements}

I am grateful to N. Sakai and A. Migdal for illuminating discussions.

\end{document}